\documentclass[aps,prb,footinbib,showkeys,showpacs,twocolumn,floatfix]{revtex4}
\usepackage[dvips]{graphicx}
\usepackage{mathrsfs}
\usepackage{amsmath}
\usepackage{amssymb}
\usepackage{amsfonts}
\usepackage{ae}
\usepackage{units}
\begin{document}
%
%-----------title---------------
\title{Voltage-driven superconducting weak link as a refrigerator for cooling of nanomechanical vibrations}
\author{Gustav Sonne}
\email{gustav.sonne@physics.gu.se}
\affiliation{\small University of Gothenburg, Department of Physics, SE-412 96 G\"oteborg, Sweden}
\author{Milton E. Pe\~na-Aza}
\affiliation{\small University of Gothenburg, Department of Physics, SE-412 96 G\"oteborg, Sweden}
\author{Leonid Y. Gorelik}
\affiliation{\small Chalmers University of Technology, Department of Applied Physics, SE-412 96 G\"oteborg, Sweden}
\author{Robert I. Shekhter}
\affiliation{\small University of Gothenburg, Department of Physics, SE-412 96 G\"oteborg, Sweden}
\author{Mats Jonson}
\affiliation{\small University of Gothenburg, Department of Physics, SE-412 96 G\"oteborg, Sweden}
\affiliation{\small Heriot-Watt University, School of Engineering and Physical Sciences, Edinburgh EH14 4AS, Scotland, UK}
\affiliation{\small Konkuk University, School of Physics, Division of Quantum Phases and Devices, 143-107 Seoul, Korea}
\date{\today}
%
%----abstract
\begin{abstract}
We consider a new type of cooling mechanism for a suspended nanowire
acting as a weak link between two superconductive electrodes. By
applying a bias voltage over the system, we show that the system can
be viewed as a refrigerator for the nanomechanical vibrations, where
energy is continuously transferred from the vibrational degrees of
freedom to the extended quasiparticle states in the leads through the
periodic modulation of the inter-Andreev level separation. The
necessary coupling between the electronic and mechanical degrees of
freedom responsible for this energy-transfer can be achieved both with
an external magnetic or electrical field, and is shown to lead to an
effective cooling of the vibrating nanowire. Using realistic
parameters for a suspended nanowire in the form of a metallic carbon
nanotube we analyze the evolution of the density matrix and
demonstrate the possibility to cool the system down to a stationary
vibron population of $\sim 0.1$. Furthermore, it is shown that the
stationary occupancy of the vibrational modes of the nanowire can be
directly probed from the DC current responsible for carrying away the
absorbed energy from the vibrating nanowire.
\end{abstract}
\maketitle
%
%\tableofcontents
\def\thesection{\arabic{section}}
\section{Introduction}
%%%%%%%%%%%%%%%%%%%%%%%%%%%%%%%%%
%%%%%%%%%%%%%%%%%%%%%%%%%%%%%%%%%
Nanoelectromechanical systems (NEMS) have over the last two decades
been a very active field of both fundamental and applied
research. These systems have typical dimensions on the nanoscale and
combine electronic and mechanical degrees of freedom for novel
applications. These include ultra sensitive mass detection
[\onlinecite{Lassagne2008,Jensen2008,Naik2009}] and position sensing
[\onlinecite{LaHaye,Etaki2008,Hertzberg2010}] using small mechanical
resonators with a resolution that has been shown to approach the limit
set by the Heisenberg uncertainty principle. Due to the small
dimensions implied by NEMS they border on the world of quantum
mechanics while still being macroscopic in the sense that they can be
fabricated lithographically. Nanoelectromechanical systems thus allow
for the controlled fabrication of devices whose physics is ultimately
governed by the fundamental limits set by quantum mechanics. Achieving
control and understanding of these systems is therefore of great
interest as doing so opens up a whole new toolbox for the design of
high-performance applications [\onlinecite{Schwab2005,Blencowe2004}].

Typically, nanoelectromechanical systems comprise a mechanical
resonator in the form of a cantilever or a doubly-clamped beam coupled
to an electronic system used for both actuation and detection. Due to
the high mechanical vibration frequencies and exceptionally high
quality factors recently achieved, these systems allow for very low
energy dissipation and extreme sensitivity to external stimuli. To
make full use of their potential much research has recently focused on
the possibility to effectively cool the mechanical subsystem in NEMS
to its vibrational ground state, which would enable unprecedented
mass- and position-sensitivity of the resonators. Recently, O'Connell
and colleagues [\onlinecite{OConnel2010}] showed that complete ground
state cooling of a mechanical resonator can be achieved at very low
temperatures. In Ref.~[\onlinecite{OConnel2010}] the reported vibron
occupancy factor is $<0.07$, which means the resonator is in its
ground state with a probability of $93\%$, well within the quantum
regime. Here, we note however, that this effect is attributed to the
high mechanical frequency of the resonator (in the GHz range) and the
very low (\unit[25]{mK}) temperature of the dilution fridge. No active
cooling of the resonator is needed as the low occupancy of its excited
vibrational modes simply follows from the Bose-Einstein distribution
function at the equilibrium temperature. To date, the best
experimental result for active cooling of a mechanical resonator lead
to a vibron occupation factor of $3.8$, as recently reported by
Rocheleau {\it et al.}  [\onlinecite{Rocheleau2009}]. In
Ref.[\onlinecite{Rocheleau2009}] an external electromagnetic field
acts as an energy transducer for side-band cooling of the mechanical
resonator (see also Ref.~[\onlinecite{Naik}], where dynamical
back-action is used to achieve similar results).

\begin{figure}
\includegraphics[width=0.45\textwidth]{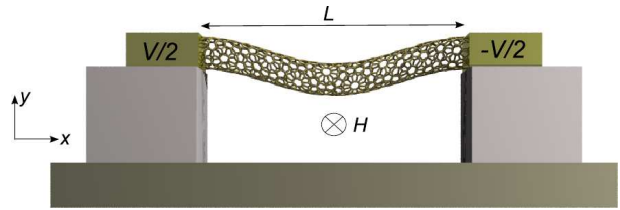}
\caption{(Color online) Schematic diagram of the system considered. A
  suspended carbon nanotube of length $L$ is coupled to two
  superconducting leads biased at a voltage $V$. Coupling between the
  Andreev states for the electronic degrees of freedom and the
  mechanical vibrations of the nanowire is achieved through the
  external transverse magnetic field $H$ which enables transition
  between the electronic branches through the emission/absorption of a
  vibrational quantum (see text).}
\label{picture}
\end{figure}

In this paper we will consider a new experimental setup for ground
state cooling of a mechanical resonator, a nanomechanical
superconducting weak link as shown schematically in
Fig.~\ref{picture}. Below we show that this system can be viewed as a
refrigerator for the nanomechanical vibrations, where the
voltage-driven Andreev states serve as the refrigerant responsible for
pumping energy from the nanomechanical vibrations to the thermostat of
quasiparticle states in the leads. The possibility of such an energy
transfer is based on the periodic, in time, modulation of the
inter-Andreev level spacing and the coupling between these current
carrying Andreev states and the mechanical degrees of freedom through
an applied magnetic field (we note that this coupling could also be
achieved through an external electric field as outlined in
Appendix~\ref{AppendEfield}). At the start of each cooling cycle the
Andreev level separation initially shrinks, which brings them into
thermal contact with the vibronic degrees of freedom, enabling energy
exchange between the two systems. During this time, the work done
by the bias voltage results in an ``over-cooling'' of the Andreev
level population, which makes the heat transfer from the
nanomechanical to the electronic degrees of freedom possible. After a
while the inter-Andreev level separation increases again, eventually
reaching its initial value, at which point the Andreev states release
their excess energy into the extended quasiparticle states of the
leads. The cooling process suggested here thus models a nanomechanical
refrigerator where heat is continuously transferred along the chain;
nanomechanical vibrations $\rightarrow$ Andreev levels $\rightarrow$
quasiparticle states, resulting in an overall effective cooling of the
nanomechanical vibrations.

Unlike many recent theoretical models in which electromechanically
induced ground state cooling of mechanical resonators have been
reported
[\onlinecite{Martin2004,Zippilli2009,WilsonRae2004,Ouyang2009,Zippilli2010}],
the cooling mechanism suggested here relies neither on the
abovementioned side-band effect nor on dynamical back-action. Rather,
the cooling is an inherent property of the device which makes the
suggested mechanism a promising candidate for truly quantum mechanical
manipulation of mechanical resonators.

The electromechanical coupling required for our cooling mechanism to
work will be assumed to be due to an external magnetic field applied
perpendicular to the length of the nanowire
[\onlinecite{Sonne2010}]. Displacements of the nanowire will then
couple the mechanical and electronic degrees of freedom through the
Lorentz force. For a ``short'' SNS junction (with a coherence length
much longer than the nanowire) the AC Josephson dynamics of the
device, induced by a weak DC driving voltage, $eV<(\Delta
E(\phi))^2/\Delta_0$, can be expressed in terms of a pair of Andreev
states, which are periodically created and destroyed in the
junction. Here, $\Delta E(\phi)$ is the distance between the Andreev
levels, $\phi=\phi(t)$ is the voltage-driven phase difference over the
junction and $\Delta_0$ is the order parameter in the superconductors.
The creation/destruction of the Andreev states takes place at the
edges of the superconducting gap, $\Delta E=2\Delta_0$, when
$\phi(t)=2\pi n$ (see Fig.~\ref{Andreevpic}). Between creation and
destruction the Andreev levels evolve adiabatically within the energy
gap of the superconductors, such that the inter-Andreev level spacing
first shrinks and is later restored to its initial value $2\Delta_0$
at a time when the Andreev states dissolve in the continuum
quasiparticle spectrum (see Fig.~\ref{Andreevpic}). The whole cycle
then repeats itself again and again.
%------------------------
%----------------------

Being thermally populated at the moment of their nucleation, the
Andreev states will initially, as the inter-level spacing decreases,
experience an effective cooling given that the thermal relaxation is
not fast enough to follow the level displacement. Under such
conditions, the Andreev states can accumulate energy from any external
system (in particular from the nanomechanical vibrations) if
interactions with such a system are established. The absorbed energy
is then transferred into the electronic quasiparticle continuum when
the Andreev states are dissolved and the process of cooling of the
external system is continued through the formation of new thermally
populated Andreev states, which in their turn evolve with time and
absorb energy. After each full period of evolution the extra energy
absorbed in this process is removed from the SNS junction by the
quasiparticle excitations.

Besides being a new type of system where efficient cooling of
nanomechanical-resonators can be achieved, this type of
superconducting weak links makes it possible to couple the
nanomechanical system of the device to the macroscopically coherent
superconducting states. It presents an exciting possibility to
entangle the current carrying Andreev states associated with the SNS
junction to the quantum states of nanomechanical vibration when the
mechanical subsystem is sufficiently cooled. Manipulation of such
nanoelectromechanical coherent states can be done both electronically,
by means of driving- and gate voltages, or mechanically, by
controlling the displacement of a movable part of the device.

\section{Model Hamiltonian}
%%%%%%%%%%%%%%%%%%%%%%%%%%%%%%
%%%%%%%%%%%%%%%%%%%%%%%%%%%%%%%
To discuss the cooling mechanism quantitatively we use the
Hamiltonian,
%--------------------
\begin{gather}
\label{Hamil}
\hat{H}(t)=\hat{H}_{el}+\hat{H}_{mech}+\hat{H}_{int}\\
\hat{H}_{el}(t)=\begin{pmatrix}-\frac{\hbar^2}{2m}\frac{\partial^2}{\partial x^2}-\mu & \Delta(x) e^{i\phi(t)}\\
\Delta(x)e^{-i\phi(t)} &\frac{\hbar^2}{2m}\frac{\partial^2}{\partial x^2}+\mu\end{pmatrix}\\
\label{HBdG}
\hat{H}_{mech}=\hbar\omega\hat{b}^{\dagger}\hat{b}\,.
\end{gather}
%---------------------
In \eqref{Hamil}, $\hat{H}_{el}(t)$ is the Bogoliubov-de Gennes (BdG)
Hamiltonian for the electronic degrees of freedom.  Here,
$\Delta(x)=\Delta_0\Theta(\vert x\vert-L/2)$ is the order parameter in
the superconductive leads (here taken to be identical), $\mu$ is the
chemical potential and $\phi(t)$, the phase difference over the
junction, depends on the bias voltage $V$ according to the Josephson
relation $\dot{\phi}=2eV/\hbar$. The second term in the Hamiltonian
describes the oscillating nanowire which is modeled as a quantum
mechanical harmonic oscillator with only the fundamental mode excited;
$\hat{b}^{\dagger}$ $[\hat{b}]$ is an operator that creates
[annihilates] a quantum of vibration and $\omega$ is the frequency of
the fundamental mode. In \eqref{Hamil}, $\hat{H}_{int}$ is the
interaction Hamiltonian which gives the coupling between the
electronic and the mechanical degrees of freedom through the magnetic
field.

At the superconductor-nanowire interface two electronic scattering
processes are possible, viz. normal reflection and Andreev
reflection. These give rise to two localized states --- Andreev states
--- described by the wavefunctions which satisfy
$\hat{H}_{el}\psi_{\pm}(x,t)=E_{\pm}\psi_{\pm}(x,t)$ where
$\psi_{\pm}(x,t)$ are two-state spinors in Nambu space, whose
eigenenergies are given by the expression
$E_{\pm}(t)=\pm\Delta_0(1-D\sin^2(\phi(t)/2))^{1/2}$, where $D=1-R$ is
the normal junction transparency
[\onlinecite{Bagwell,Beenakker1991}]. These Andreev states are
responsible for carrying the Josephson supercurrent through the
(normal) nanowire. Under the condition that these states evolve slowly
in time ($\hbar\dot{\phi}\ll \Delta_0$), they describe a two-level
system for the electronic degrees of freedom. As the Andreev states
carry current through the oscillating nanowire, they will couple the
electronic degrees of freedom to the mechanical degrees of freedom
through the Lorentz force. It is thus more natural to describe the
system in terms of the temporally evolving Andreev states, to which
purpose we rewrite the Hamiltonian as
%---------------------
\begin{gather}
\hat{H}(t)=\mathscr{E}(\phi(t))\hat{\sigma}_z+\Delta_0\sqrt{R}\sin(\phi(t)/2)\hat{\sigma}_x+\notag\\
\hbar\omega\hat{b}^{\dagger}\hat{b}+\frac{2e}{\hbar}\frac{\partial \mathscr{E}(\phi(t))}{\partial \phi}LH\hat{y}\hat{\sigma}_z\,.
\label{Hamilnew}
\end{gather}
%----------------
Here, $\mathscr{E}(\phi(t))$ is the energy of the Andreev states for
the completely transparent ($R=0$) junction and $\hat{\sigma}_i$
($i=x,y,z$) are the Pauli matrices. The last term in \eqref{Hamilnew}
describes the electromechanical coupling of the current-carrying
Andreev states to the motion of the nanowire through the Lorentz
force. Here, $e$ is the electronic charge, $L$ is the length of the
wire, $\hat{y}=y_0(\hat{b}^{\dagger}+\hat{b})$ --- where $y_0$ is the
zero-point oscillation amplitude --- is the deflection operator in the
transverse $y$-direction, $H$ is the magnetic field and $2e/\hbar(\partial
\mathscr{E}/\partial\phi)\hat{\sigma}_z$ is the current operator in
the nanowire [\onlinecite{Gorelik1995}].
%----------------
\begin{figure}
\includegraphics[width=0.4\textwidth]{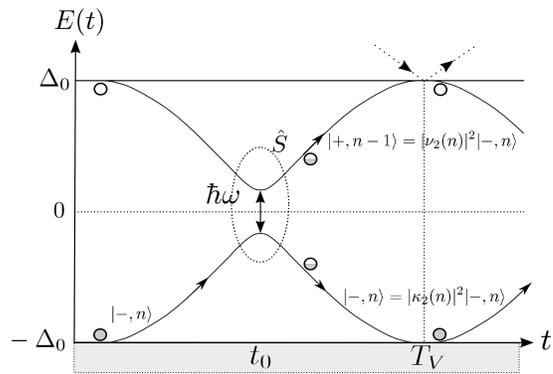}
\caption{Time evolution of the Andreev states (full lines) over one
  period $T_V=\pi\hbar/eV$. The state of the total system,
  $\vert\pm,n\rangle$, depends on the population of the two electronic
  branches, $\vert\pm\rangle$, corresponding to the upper and lower
  branch respectively, as well as on the quantum state of the
  oscillating nanowire, $\vert n\rangle$. Due to the large separation
  in energies between $\Delta_0$ and $\hbar\omega$, transitions
  between the electronic branches is only possible in the small
  resonance-window $t_0-\delta t\leq t\leq t_0+\delta t$ (modeled
  through the scattering matrix $\hat{S}$) where the electronic state
  of the system can change through the emission/absorption of one
  vibrational quantum. In the above, the probability for the state
  initially in $\vert -,n\rangle$ to scatter into the state $\vert
  +,n-1\rangle$ after passing through the resonance depends on the
  state of the oscillator through the coefficient
  $\vert\nu_2(n)\vert^2$ (see text). After one period the partially
  filled Andreev levels join the continuum, a process which is here
  represented by dashed arrows, and the electronic
  states are reset (filled and empty circles).}
\label{Andreevpic}
\end{figure}
%-------------------

From Fig.~\ref{Andreevpic} one can see that the cooling mechanism
described is maximally efficient if the strongest coupling of the
mechanical subsystem to the Andreev states occur during the time
interval when the Andreev levels are closest to the chemical
potential, i.e., when the population of the Andreev levels differ most
from what it would be in thermodynamic equilibrium. The coupling is
resonant during this time if $\hbar\omega=\textrm{min}(\Delta
E(\phi))=2\Delta_0\sqrt{R}$ (where $R=1-D$ is the normal-state
reflection probability), which requires the weak link to have a high
normal-state transparency $D$ for electrons
[\onlinecite{Kong2001}]. This is because, for the system considered,
the energy-scales of the superconductive order parameter and the
mechanical vibrations, $\hbar\omega$, are very different. Also, one
notes that when $\phi=\pi$, the Andreev states $\psi_{\pm}(\phi=\pi)$,
with energies $E_{\pm}(\phi=\pi)=\pm\Delta_0R^{1/2}$, are
symmetric/antisymmetric superpositions of states carrying current in
opposite directions. Transitions between the Andreev states ---
induced by the nanowire as it vibrates in a transverse magnetic field
--- are therefore by far most probable when $\phi(t=t_0)=\pi$.

In the adiabatic regime, i.e. when $eV\leq eV_c=4R\Delta_0$, it is
convenient to switch to the basis set formed by the Andreev states
$\psi_{\pm}(\phi(t))$\footnote{In this regime, the
  voltage-induced Landau-Zener transition at $\phi=\pi$ between the
  adiabatic Andreev levels is suppressed, ensuring that without
  electromechanical coupling the level population remains constant in
  time.}. In this basis the Hamiltonian reads,
%--------------------
\begin{align}
\label{finalH}
\hat{H}_{eff}(t)=\langle \psi_{\pm}\vert\hat{H}(t)&\vert\psi_{\pm}\rangle= E(t)\hat{\tau}_z+\notag\\
&\hbar\omega\hat{b}^{\dagger}\hat{b}+\Delta_0\Phi(\hat{b}^{\dagger}+\hat{b})\hat{\tau}_x\,.
\end{align}
%-----------------------
Here, the Pauli matrices $\hat{\tau}_i$ span the space formed by the
states $\psi_{\pm}(\phi)$. In the last term of \eqref{finalH},
$\Phi=2LH\pi y_0/\Phi_0$ is the dimensionless magnetic flux threading
the area swept by the nanowire as it vibrates in the ground state;
$\Phi_0=h/2e$ is the magnetic flux quantum. Below we will consider the
resonant situation when $\hbar\omega=2\Delta_0R^{1/2}$, which is the
optimum condition for the proposed cooling mechanism as outlined
above\footnote{This condition can be achieved by controlling the
  magnitude of the order parameter $\Delta_0$ through the magnetic
  field for junctions with small reflection coefficients.}. 

The Hamiltonian \eqref{finalH} describes the evolution of the Andreev
states. Coupling between these is achieved through the off-diagonal
terms in \eqref{finalH}. These are proportional to $\Phi$ and become
relevant whenever the energy separation between the electronic
branches matches the energy scale set by the oscillating nanowire. At
such times, transitions between the Andreev states can occur
through the absorption/emission of one vibrational quantum. This is
shown in Fig.~\ref{Andreevpic} for the case when only the lowest
electronic branch is initially populated.

\section{Magnetic field induced Andreev state scattering}
%%%%%%%%%%%%%%%%%%%%%%%%%%%%%%%%%%%%%%%%%%
%%%%%%%%%%%%%%%%%%%%%%%%%%%%%%%%%%%%%%%%%
Assuming the conditions for adiabaticity outlined above
fulfilled, we now evaluate the probability of mechanically induced
Andreev level transitions as a function of the coupling strength. Due
to the large separation of energy scales, $2\Delta_0\gg \hbar\omega$,
it is not possible to evaluate such transitions using the Landau-Zener
formalism [\onlinecite{Gorelik1995,Gorelik1998}] as this assumes a
linear expansion of $E(t)$ around the resonance point. Rather, we
consider transitions only to occur in the vicinity of
$t_0=T_V/2=\pi\hbar/2eV$ and evaluate \eqref{finalH} using a parabolic
expansion of $E(t)$. From \eqref{finalH}, the dimensionless
time-dependent Schr\"{o}dinger equation for the probability amplitudes
$c_{\pm,n}(\tau)$ of finding the state of the system in the
upper/lower electronic branch with the oscillator in the state $n$
reads,
%---------------------
\begin{widetext}
\begin{subequations}\label{probs}
\begin{gather}
\label{probampl1}
i\partial_\tau c_{+,n-1}(\tau)=\tau^2c_{+,n-1}(\tau)+\Gamma\left(\sqrt{n}c_{-,n}(\tau)+\sqrt{n-1}e^{2i\lambda\tau}c_{-,n-2}(\tau)\right)\\
i\partial_\tau c_{-,n}(\tau)=-\tau^2c_{-,n}(\tau)+\Gamma\left(\sqrt{n}c_{+,n-1}(\tau)+\sqrt{n+1}e^{-2i\lambda\tau}c_{+,n+1}(\tau)\right)
\label{probampl2}
\end{gather}
\end{subequations}
\begin{gather}
\tau=(t-t_0)\left(\frac{\xi}{\hbar}\right)^{1/3} \qquad \Gamma=\frac{\Phi\Delta_0}{\hbar\omega}\left(\frac{V_c}{V}\right)^{2/3}\qquad
\xi=\frac{\partial^2E(t)}{\partial t^2}\bigg\vert_{t_0}=\frac{D(\hbar\omega)^3}{\hbar^2}\left(\frac{V}{V_c}\right)^2\qquad
\lambda \tau=\tau\left(\frac{V_c}{V}\right)^{2/3}+\frac{\pi\hbar\omega}{2eV}\notag\,.
\end{gather}
\end{widetext}
%----------------------

Equation \eqref{probs} describes the adiabatic evolution of the
electronic Andreev states. As theses states pass through the resonance
they couple to the mechanical subsystem, enabling transitions between
them through the emission/absorption of a quantum of vibration as
indicated by the terms proportional to the dimensionless
electron-vibron coupling constant $\Gamma$ in
\eqref{probs}. Considering for now the situation when the system
initially starts in the lower electronic branch \eqref{probampl1} we
analyze the probability that the system, after passing through $t_0$,
scatters into the upper electronic branch as a function of the
coupling strength. As can be seen from \eqref{probampl1}, two
scattering processes are possible; absorption of a mechanical vibron
(proportional to $\sqrt{n}$) or emission of a vibron (proportional to
$\sqrt{n-1}$). Note however, that these two processes differ by an
exponential phase factor associated with the emission process.

Considering that the coupling between the electronic and mechanical
degrees of freedom is very weak (see below) we evaluate the
probability for the electronic sub-system to scatter into the upper
electronic branch perturbatively in the small parameter $\Gamma$. From
this analysis we find that scattering is much more likely to occur
through the absorption process than through the emission process,
whose importance is reduced by the exponential phase factor mentioned
above. A numerical analysis shows that the probability of
magnetic-field induced Andreev state scattering through the emission
channel is only $\sim 3\%$ of the probability of scattering through
the absorption channel. For the case when the system is initially in
the upper electronic state \eqref{probampl2} the same analysis shows
that the emission process (proportional to $\sqrt{n}$) is much more
likely than the absorption process (proportional to
$\sqrt{n+1}$). Thus, we can safely treat the absorption process (when
the electronic system is initially in the lower electronic branch) as
the most dominant process, i.e. we may apply the rotating wave
approximation (RWA) to \eqref{probs}. Note, however, that without
thermal damping of the nanowire vibrations caused by the external
environment (see below), the rate of emission would ultimately limit
the efficiency of our cooling process as it would prohibit complete
ground state cooling. The details of such an analysis is presented in
Appendix~\ref{AppendRWA}, where we show that by not using the rotating
wave approximation the lowest theoretically achievable level of
cooling of the nanowire corresponds to a population of the vibrational
mode of $\langle n\rangle\simeq 0.03$.

Within the rotating wave approximation we proceed to evaluate the
scattering probability as a function of the coupling strength in more
detail, only keeping the terms proportional to $\sqrt{n}$ in
\eqref{probs},
%-----------------
\begin{gather}
\label{finalprobampl}
i\partial_{\tau}\vec{c}_n(\tau)=\left(\tau^2\tau_z+\Gamma\sqrt{n}\tau_x\right)\vec{c}_n(\tau)\\
\vec{c}_n(\tau)=\left(c_{+,n-1}(\tau),c_{-,n}(\tau)\right)^{T}\,.\notag
\end{gather}
%------------------

The electronic scattering mechanism considered here is similar to the
Landau-Zener formalism in that the probability of scattering depends
on the ratio between the coupling strength and the rate of evolution
through the interaction region, off-diagonal terms in
\eqref{finalprobampl}. However, unlike in the Landau-Zener approach,
where in the corresponding equation the dependence on $\tau$ is only
linear, equation \eqref{finalprobampl} cannot be solved
analytically. Instead we have performed numerical simulations in order
to find the probability of scattering as a function of the coupling
strength. From this analysis we find that unlike the Landau-Zener
formula, which gives a unitary transition probability in the strong
coupling limit, the parabolic dependence on $\tau$ in
\eqref{finalprobampl} dictates that the probability of transition in
the limit of strong coupling (or infinitely slow transition through
the interaction region) is only $0.5$. Thus, for very low applied bias
voltages we should expect a probability of transition between the
Andreev states of $0.5$ each time the system passes through the
resonance.

In the adiabatic regime, $V\lesssim V_c$ the coupling for realistic
experimental parameters is, however, relatively weak;
$\Gamma=\Delta_0\Phi/(\hbar\omega)\ll 1$, hence we can analyze
\eqref{finalprobampl} treating $\Gamma$ as a perturbation to the
solution $\vert c_{-,n}(-\delta \tau)\vert^2=1$, where the system
starts in the lower electronic branch at time $-\delta\tau=-\delta
t(\xi/\hbar)^{1/3}$. Here, $\delta
t\simeq(\hbar\omega\Phi/\Delta_0)^{1/2}\hbar/(eV)\ll \pi\hbar/(eV)$ is
the characteristic time scale of the electromechanical
interactions. Solving \eqref{finalprobampl} under this condition for
the probability $\vert c_{+,n-1}(\delta \tau)\vert^2$ of the system to
be in the upper branch after passing through the resonance we find
$\vert c_{+,n-1}(\delta\tau)\vert^2\simeq n\pi \Gamma^2$ which is
consistent with our numerical solutions.  Thus, for weak coupling
between the electronic and mechanical degrees of freedom, the
probability for a transition between the Andreev states to have
occurred after passing through the resonance (with the corresponding
absorption of a mechanical vibron) scales linearly with the initial
population of the mechanical resonator. Furthermore, as can be seen
from the symmetry of \eqref{finalprobampl}, this expression also gives
the probability of finding the system in the state $\vert -,n\rangle$
given that it was nucleated in $\vert+,n-1\rangle$.

\section{ground state cooling of oscillating nanowire}
%%%%%%%%%%%%%%%%%%%%%%%%%%%%%%%%%%%%%%%%%%%%%%%%%
%%%%%%%%%%%%%%%%%%%%%%%%%%%%%%%%%%%%%%%%%%%%%%%%%
In order to model the full evolution of the coupled electromechanical
system we evaluate the total density matrix $\hat{\rho}$ of the system
over one period. Under the assumption that the thermal energy at the
ambient temperature, $T$, is small compared to the initial separation
between the electronic branches, $2 \Delta_0/k_B T>1$, the system
will, at the start of the period, be found in the lower electronic
branch with a distribution, $P_n^{in}$, of the mechanical excitations,
%----------------------
\begin{equation}
\hat{\rho}^{in}=\sum_{i,j=\pm}\sum_{n=0}^{\infty}P_n^{in}\vert i,n\rangle\langle j,n\vert=\sum_{n=0}^{\infty}P_n^{in}\begin{pmatrix}0 &0\\0 &1\end{pmatrix}\vert n\rangle \langle n\vert\notag\,.
\end{equation}
%-----------------------
During the adiabatic evolution (no coupling between branches) the
system interacts with the external heat bath and the rate of change of
the density matrix is given by,
%-------------------
\begin{align}
\frac{\partial \hat{\rho}(t)}{\partial t}=-\frac{i}{\hbar}&\left[\hat{H}_{eff}(t),\hat{\rho}(t)\right]+\frac{\gamma}{2}\hat{\mathfrak{L}}(\hat{\rho}(t))\,,
\end{align}
%---------------
where the collision integral
%-------------------------
\begin{align}
\hat{\mathfrak{L}}(\hat{\rho})=-(1&+n_B)\left(\hat{b}^{\dagger}\hat{b}\hat{\rho}+\hat{\rho}\hat{b}^{\dagger}\hat{b}-2\hat{b}\hat{\rho}\hat{b}^{\dagger}\right)-\notag\\
&n_B\left(\hat{b}\hat{b}^{\dagger}\hat{\rho}+\hat{\rho}\hat{b}\hat{b}^{\dagger}-2\hat{b}^{\dagger}\hat{\rho}\hat{b}\right)\,,
\end{align}
%----------------------------
models the interaction of the mechanical subsystem with the
environment. Here, $n_B=(\exp(\hbar\omega/k_BT)-1)^{-1}$ and
$\gamma=\omega/Q$ is the thermal damping rate of the vibrational modes
with $Q$ the quality factor. Considering the coupling to the
environment to be small ($Q\sim$\unit[$10^{5}$]{}
[\onlinecite{Huttel2009}]) we can solve for the evolution of the
density matrix by treating the interaction of the system with the
external environment as a perturbation.

The adiabatic evolution of the electronic system ensures that the
coupling between the Andreev states is virtually zero. Only in the
small region around $t_0$ do the two branches interact (through the
coupling to the mechanical degrees of freedom), which is conveniently
accounted for by describing the evolution of $\hat{\rho}$ through the
resonance by the unitary scattering matrix $\hat{S}$,
%----------------------
\begin{gather}
\label{sth}
\hat{\rho}(t_0+\delta t)=\hat{S}\hat{\rho}(t_0-\delta t)\hat{S}^{\dagger}\\
\label{Smatrix}
\hat{S}=\begin{pmatrix}\kappa_1(\hat{n})& i\frac{\nu_1(\hat{n})}{\sqrt{\hat{n}+1}}\hat{b}\\ i\hat{b}^{\dagger}\frac{\nu_2(\hat{n}+1)}{\sqrt{\hat{n}+1}} &\kappa_2(\hat{n})\end{pmatrix}\\
\vert\kappa_i(n)\vert^2+\vert\nu_i(n)\vert^2=1\notag\,.
\end{gather}
%----------------------
Here, $\hat{n}=\hat{b}^{\dagger}\hat{b}$ is the vibron number operator
and the subscripts $1,2$ refer to the top/bottom Andreev state
respectively. The coefficients $\kappa_i(\hat{n})$ [$\nu_i(\hat{n})$]
are respectively the probability amplitude for the system to stay
[scatter] out of the initial Andreev state $i$, both of which depend
on the state of the oscillator as outlined above. As such,
$\vert\nu_2(n)\vert^2$ is the probability of the system, initially in
the state $n$ in the lower electronic branch, to scatter into the
upper electronic branch through the absorption of a vibron. It thus
corresponds to $\vert c_{+,n-1}(\delta \tau)\vert^2$ from
\eqref{finalprobampl} with the initial state $\vert -,n\rangle$. With
this we have $\vert\nu_2(n)\vert^2\simeq \pi n\Gamma^2$ and also
$\vert\nu_2(n)\vert^2=\vert\nu_1(n-1)\vert^2$ which follows from the
symmetry of \eqref{finalprobampl}.

After one period the partially filled Andreev branches merge with the
superconducting continuum. Here, non-adiabatic interactions release
the energy of the charge-carrying quasiparticles into the continuum
and the initial conditions for the Andreev level population is reset
[\onlinecite{Gorelik1998}], i.e. the electronic system returns to the
lower branch (see Fig.~\ref{Andreevpic}). The mechanical system, which
forms the major interest of this paper, will however not return to the
initial distribution after one period. Thus, we may find the
density matrix for the mechanical system after one period,
$\hat{\rho}_{mech}^{f}$, by tracing out the electronic degrees of
freedom of the total density matrix,
%-------------------
\begin{align}
\hat{\rho}^{f}=&\sum_{n=0}^{\infty}P_n^{f}\begin{pmatrix}0 &0\\0 &1\end{pmatrix}\vert n\rangle \langle n\vert\notag\\
\label{rhoevol}
\hat{\rho}_{mech}^{f}=\textrm{Tr}_{el}&(\hat{\rho}^{f})=\textrm{Tr}_{el}\bigg(\hat{S}\hat{\rho}^{in}\hat{S}^{\dagger}+\notag\\
\frac{\gamma}{2}&\frac{T_V}{2}\hat{S}\hat{\mathfrak{L}}(\hat{\rho}^{in})\hat{S}^{\dagger}+\frac{\gamma}{2}\frac{T_V}{2}\hat{\mathfrak{L}}(\hat{S}\hat{\rho}^{in}\hat{S}^{\dagger})\bigg)\,.\end{align}
%-----------------
%-------------------
\begin{figure}
\includegraphics[width=0.4\textwidth]{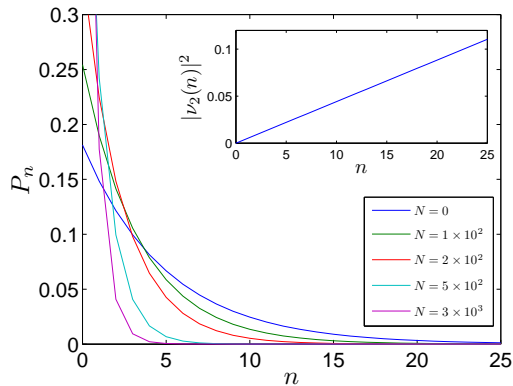}
\caption{(Color online) Evolution of the distribution of the
  mechanical modes, $P_n$, as a function of the quantum state $n$ for
  different number of periods $N$
  ($T_V=\pi\hbar/eV\sim$\unit[20]{ns}). Initially $P_n$ is thermally
  distributed $P_n\propto e^{-\hbar\omega n/k_BT}$ with
  $k_BT=5\hbar\omega$. Here, $V=$\unit[$10^{-7}$]{V},
  $\hbar\omega=$\unit[$10^{-6}$]{eV}, $\Delta_0=10\hbar\omega$,
  $y_0=$\unit[20]{pm}, $L=$\unit[100]{nm} and $H=$\unit[1]{T}. Inset
  shows $\vert\nu_2(n)\vert^2$ as a function of $n$ for the same
  parameters.}
\label{densitypic}
\end{figure}
%--------------------------

Equation \eqref{rhoevol} describes the evolution of the mechanical
density matrix over one period. Without coupling to the environment,
$\gamma=0$, this corresponds to the decay of the mechanical subsystem
as for each period there is a probability $\vert\nu_2(n)\vert^2$ that
the state $\vert -,n\rangle$ scatters into the state $\vert
+,n-1\rangle$. As the opposite process is forbidden if initially only
the lower electronic branch is populated, the mechanical subsystem
would thus approach the vibrational ground state at a rate which
depends on the strength of the coupling. As noted above, complete
ground state cooling would however not be possible due to the small
but finite probability of inter-Andreev level scattering through the
emission channel.

When including effects of the environment, the mechanical subsystem
does not necessarily decay to the ground state since the external
damping drives the system towards thermal equilibrium. In
Fig.~\ref{densitypic} we plot the evolution of the distribution of the
vibrational states, $P_n$, as a function of the quantum numbers $n$
over many periods for realistic experimental parameters. As can
clearly be seen, the above discussed scattering process acts to lower
the initially thermally distributed population of the mechanical
system, effectively cooling the nanowire down to a final population of
the vibronic states of $\langle n\rangle =\sum_nP_nn\sim 0.1 $. Note,
however, that complete ground state cooling is not achieved due to the
temperature of the external environment.

To evaluate the efficiency of the discussed cooling mechanism we
analyze the stationary solution of \eqref{rhoevol}, i.e. when the
mechanical subsystem does not change over one period, perturbatively
in the small parameter $\gamma$. From this analysis we find that to
zeroth order in $\gamma$ the stationary solution gives
$\vert\nu_2(n)\vert^2P_n=0$ which corresponds to the distribution of
the mechanical modes decaying to the ground state
($\vert\nu_2(0)\vert^2=0$). The corrections to this solution due to
the external heat bath is found from the higher order terms in the
pertubative expansion where the lowest order correction to the state
$n$ scales as $\gamma^n$. We thus find that to linear order in
$\gamma$ the correction to the distribution of the mechanical modes in
the stationary regime, $P_n^{stat}$, is given by
%-------------------
\begin{gather}
\label{rhostat}
P_n^{stat}=P_n^{0}+\gamma\frac{\pi\hbar n_B}{2eV}\delta P_n+O(\gamma^2)\\
P_n^{0}=P_0^0\delta_{n,0}\notag\\
\delta P_n=(-1)^{n+1}\frac{(2-\vert\nu_2(1)\vert^2)}{\vert\nu_2(1)\vert^2}\qquad n=0,1\notag\,.
\end{gather}
%-----------------------
With this, we find that the population of the first excited state in
the stationary regime can be expressed as,
%-----------------
\begin{equation}
P_1^{stat}\simeq\frac{n_B}{Q\Phi^2}\frac{\hbar\omega}{\Delta_0}\left(\frac{V}{V_c}\right)^{1/3}\sim 0.1\,,
\end{equation}
%-------------------
which is in accordance with the results shown in
Fig.~\ref{densitypic}.

In this analysis, the strength of the electromechanical coupling,
$\Gamma$, plays two roles. First, it dictates the rate of the cooling
process as $\vert\nu_2(n)\vert^2\propto\Gamma^2$. Secondly, stronger
coupling ensures a lower final distribution of the mechanical modes;
in effect the ratio $\gamma/\Gamma^2$ dictates how many terms in the
perturbation expansion \eqref{rhostat} need to be considered in order
to accurately describe the stationary distribution of the mechanical
modes. Thus, for the parameters considered, we find that we need to
include higher order corrections in $\gamma$ in the expansion
\eqref{rhostat} to fully describe the stationary distribution. This is
confirmed in Fig.~\ref{densitypic} which shows a finite probability of
not only the ground state and first excited state being populated in
the stationary regime. Nevertheless, the expectation value of the
population of the mechanical subsystem in Fig.~\ref{densitypic} is
$\langle n\rangle\simeq 0.1$, which is well in the quantum regime.

\section{DC current measurement as a probe for the stationary vibronic population}
\begin{figure}
\includegraphics[width=0.45\textwidth]{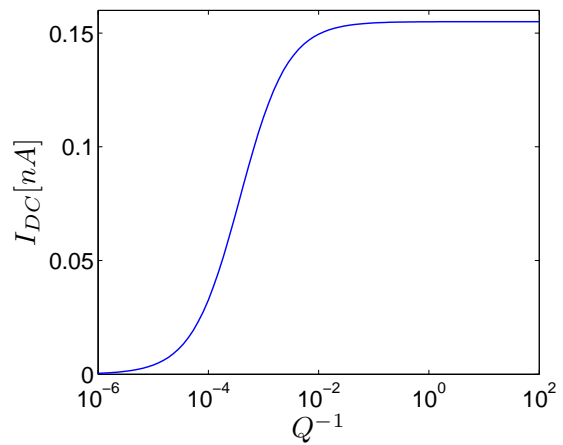}
\caption{DC current as a function of the inverse quality factor when
  the mechanical subsystem has been driven into the stationary regime.
  As can be seen, in the limit of very high quality factor,
  corresponding to complete ground state cooling of the oscillating
  nanowire, the DC current goes zero as no inter-Andreev level
  scattering is possible in this regime ($\vert\nu_2(0)\vert^2=0$). In
  the opposite limit, $Q\lesssim 10$, the DC current approaches a
  constant value which depends on the external temperature,
  $I_{DC}^{therm}\propto n_B$. Here, the system parameters are the
  same as in Fig.~\ref{densitypic}.}
\label{fig:current}
\end{figure}

%%%%%%%%%%%%%%%%%%%%%%%%%%%%%
%%%%%%%%%%%%%%%%%%%%%%%%%%%%%%
To probe the stationary distribution of the mechanical subsystem we
suggest studying the DC current through the weak link as this directly
measures the vibron population as will be shown below. By changing the
quality factor of the system, for example by varying the external
pressure, we find that the stationary vibron population can be
directly measured since it scales linearly with the DC current. To
show this we evaluate the DC current, on resonance, induced by the
inter-Andreev level scattering over one period when the mechanical
subsystem has been driven into the stationary regime. This current
comes about as the two Andreev levels carry current in opposite
directions. As such, any scattering event which populates the upper
Andreev level will result in a net charge transfer through the
system. Evaluating the total current transfer over a full period we
have,
%--------------------------------------
\begin{align}
&I_{DC}=\frac{2e}{\hbar}\frac{2}{T_V}\bigg[\int_0^{T_V/2}\frac{\partial E_-(\phi)}{\partial \phi}\textrm{d}t+\notag\\
&(1-P)\int_{T_V/2}^{T_V}\frac{\partial E_-(\phi)}{\partial \phi}\textrm{d}t+P\int_{T_V/2}^{T_V}\frac{\partial E_+(\phi)}{\partial \phi}\textrm{d}t\bigg]\,,
\end{align}
%---------------
where, $P$ is the total probability for the system to be scattered
from the lower to the upper Andreev branch during one period of
evolution. Since $n$ vibration quanta are excited with probability
$P_n^{stat}$, and since in this case the scattering probability is
$\vert\nu_2(n)\vert^2$, the total scattering probability $P$ is
readily obtained, and the expression for $I_{DC}$ simplifies to (note
that the time integration is trivial since
$\partial/\partial\phi=(\partial t/\partial \phi)\partial/\partial
t=(T_{v}/2\pi)\partial/\partial t$)
%-------------------
\begin{align}
\label{current}
I_{DC}&=\frac{4e\Delta_0}{\pi\hbar}\left(1-\frac{\hbar\omega}{2\Delta_0}\right)\sum_{n=0}^{\infty}\vert\nu_2(n)\vert^2P_n^{stat}\notag\\
&=e\langle n\rangle\Gamma^2\frac{eV_c}{\hbar}\left(\frac{1-\sqrt{R}}{R}\right)\,, 
\end{align}
%-----------------
which is plotted in Fig.~\ref{fig:current} for the same experimental
parameters as in Fig.~\ref{densitypic}.

Equation \eqref{current} shows that the DC current over the junction
scales linearly with the average vibron population in the stationary
regime. Here, we note that the correct expression for the current
should also take into account corrections from the environment as
discussed in Appendix~\ref{Appendcurr}. Numerical simulations show,
however, that the inclusion of these terms does not change the
calculated DC current as compared to that shown in
Fig.~\ref{fig:current}.

From Fig.~\ref{fig:current} we note that in the limit of small quality
factor, i.e., high damping, the DC current saturates to a constant
values.  This can be understood by the fact that in this limit the
stationary distribution of the vibrational modes is simply given by
the thermal distribution $P_n^{stat}\propto \exp(-n\hbar\omega/k_BT)$,
i.e. the interactions with the environment are so strong that they
always drive the mechanical system into thermal equilibrium, no matter
how the inter-Andreev level scattering changes this distribution. As
such, $I_{DC}\propto n_B$ in this regime. In the opposite regime,
$Q\rightarrow \infty$, the DC current goes to zero (within the
rotating wave approximation) as expected as this regime corresponds to
complete ground state cooling of the mechanical subsystem. If this can
be achieved, the probability of Andreev level scattering also goes to
zero, $\vert\nu_2(0)\vert^2=0$, hence the electronic subsystem stays
in the lower Andreev branch throughout and the $I_{DC}=0$.

From the dependency of $I_{DC}$ on $1/Q$ in Fig.~\ref{fig:current} we
suggest a simple probe by which one could measure the average vibron
population once the mechanical subsystem has been driven into the
stationary regime. We propose to change the quality factor, for
example by increasing the external pressure of the gas in which the
nanowire vibrates, and measure the DC current in the limit of large
damping. As $I_{DC}$ scales with the thermal equilibrium phonon
population, $n_B$, in this regime a further measurement in the low
damping regime should thus give a direct measurement of the average
vibron population,
%------------
\begin{equation}
\langle n\rangle=\frac{I_{DC}}{I_{DC}^{therm}} n_B\,,
\end{equation}
%---------------
where $I_{DC}^{therm}\propto n_B$ is the DC current in the limit of
high damping.
%-------------------
%--------------------------

\section{Conclusions}
%%%%%%%%%%%%%%%%%%%%%%%%%%%%%%%%%%%
%%%%%%%%%%%%%%%%%%%%%%%%%%%%%%%%%
To conclude, we have shown that quantum mechanical cooling of a
nanomechanical resonator acting as a weak link between two
superconductors is possible. In particular, we have considered the
example of a suspended carbon nanotube where the unique combination of
high resonance frequencies and mechanical quality factors together
with high electronic transparency lead to sufficiently strong coupling
for efficient cooling to be possible. Using realistic experimental
parameters we have shown that for a short suspended nanotube,
stationary distribution of the population of the vibrational modes can
be as low as $\langle n\rangle=0.1 $ which is truly in the quantum
mechanical regime. Furthermore, the suggested mechanism does not rely
on any external electromagnetic fields to stimulate the cooling
process. Rather, the proposed system should act as a self-cooling
device, given the right experimental parameters, as the over-cooled
Andreev states can readily absorb energy from the mechanical subsystem
given that sufficient coupling between the two can be achieved. The
corresponding energy uptake of the electronic subsystem is later
released into the quasiparticle continuum, leading to an effective
cooling of the nanomechanical resonator.

Furthermore, it has been shown that by measuring the DC current as a
function of the quality factor, the stationary distribution of the
population of the vibrational modes can be directly probed. Thus, the
suggested system not only has the capacity of self-cooling, but it
also presents a direct, internal probe by which to measure the
efficiency of the cooling mechanism.

%\vspace{0.5mm}{\bf Acknowledgments}
\section*{Acknowledgments}
The authors would like to thank Yury Tarakanov for help with
graphical illustrations. This work was supported in part by the
Swedish VR and SSF, by the EC project QNEMS (FP7-ICT-233952) and by
the Korean WCU program funded by MEST through KOSEF
(R31-2008-000-10057-0).

\appendix
\section{\label{AppendEfield} Electrical field induced cooling of suspended nanowire}
%%%%%%%%%%%%%%%%%%%%%%%%%%%%%%%%%%%%%%%%%%%%%%%%
%%%%%%%%%%%%%%%%%%%%%%%%%%%%%%%%%%%%%%%%%%%%%%%%
As mentioned above the required coupling for effective cooling of our
suspended vibrating nanowire could also be achieved using a transverse
electric field. Under such situations, the non-superconducting
suspended wire supplies a coupling of the Josephson current flow to
the nanomechanical bending vibrations through the displacement of the
wire with the electromechanical coupling depending on the electronic
backscattering induced by the electrical field. With this setup, the
interaction term in the Hamiltonian \eqref{Hamil} reads,
%-------------------
\begin{equation}
\hat{H}_{int}=e\mathscr{E}y_0(\hat{b}^{\dagger}+\hat{b})\hat{\sigma}_z\,,
\end{equation}
%---------------------
where $\mathscr{E}$ is the transverse electric field. Proceeding as
above, we project out the Andreev states $\psi_{\pm}$, which gives the
effective Hamiltonian,
%--------------------
\begin{align}
\hat{H}_{eff}(t)=\langle \psi_{\pm}\vert\hat{H}(t)\vert\psi_{\pm}&\rangle=\Delta E(t)\hat{\tau}_z+\hbar\omega\hat{b}^{\dagger}\hat{b}+\notag\\
&\Xi(\hat{b}^{\dagger}+\hat{b})\left(\hat{\tau}_+V_{+-}+\textrm{h.c.}\right)\\
\Delta E(t)=E(t)+\lambda(\hat{b}^{\dagger}&+\hat{b})V_{++}\qquad \Xi=e\mathscr{E}y_0\notag\,.
\end{align}
%-----------------------
In the above,
$V_{i,j}=\langle\psi_i\vert\hat{\sigma}_z\vert\psi_j\rangle$,
$\hat{\tau}_+=(\hat{\tau}_x+i\hat{\tau}_y)/2$ and $\vert
V_{+-}\vert\propto\sqrt{R}$ [\onlinecite{Gorelik1998}]. To evaluate
this further we perform the unitary transformation
$\hat{U}\hat{H}_{eff}(t)\hat{U}^{\dagger}$,
%-------------------------
\begin{gather}
\hat{U}=\exp(\alpha(\hat{b}^{\dagger}-\hat{b})\hat{\tau}_z)\qquad
\alpha=\Xi V_{++}/\hbar\omega\ll 1\,,
\end{gather}
%-------------------------
after which the Hamiltonian reads,
%------------------
\begin{align}
\label{Hfinal2}
\hat{\mathscr{H}}_{eff}(t)&=E(t)\hat{\tau}_z+\hbar\omega\hat{b}^{\dagger}\hat{b}+\notag\\
&\Xi(\hat{b}^{\dagger}+\hat{b})\left(V_{+-}\hat{\tau}_++\textrm{h.c.}\right)+O(\alpha^2)\,.
\end{align}
%-------------------------
From here on, the analysis follows the previous one with the coupling
constant now being, $\Gamma=(\Xi\vert
V_{+-}\vert/\hbar\omega)(V_c/V)^{2/3}$. For realistic experimental
parameters, our analysis shows that the coupling constant is again
much smaller than unity. Hence we can develop the same perturbative
analysis for the evolution of the density matrix. Using the same
physical parameters as above for the dimensions of the nanowire, the
external temperature and the superconductive gap with
$\mathscr{E}=$\unit[4$\times 10^4$]{Vm$^{-1}$} we find a stationary
occupancy of the vibrational modes of the oscillating nanowire of
$\langle n \rangle\sim 0.25$. Thus, this analysis shows that an
alternative approach to that followed in the main part of the paper
would be to use an electric field to provide the electromechanical
coupling. Both modes of operation give an efficient cooling of the
vibrating nanowire. However, our analysis shows that the best mode of
operation for ground state cooling should be to consider using a
magnetic field. This mode also has the added benefit that it makes it
possible to tune the superconductive order parameter with the magnetic
field to achieve resonant conditions.

\section{\label{AppendRWA} Emission limited cooling of the nanowire vibrations}
%%%%%%%%%%%%%%%%%%%%%%%%%%%%%%%%%%%%%%%%%%%%%%%%
%%%%%%%%%%%%%%%%%%%%%%%%%%%%%%%%%%%%%%%%%%%%%%%%
The analysis presented in the main part of the paper, performed within
the rotating wave approximation (RWA), showed that the suggested
cooling mechanism has the capacity of effectively reducing the
excitations of the nanowire vibrational modes well into the quantum
regime. Within the RWA, complete ground state cooling is, in theory,
achievable as a large enough quality factor limits the influence of
the environment (heating) on the mechanical sub-system. Thus, in the
limit $Q\rightarrow \infty$, thermal heating is negligible and the
inter-Andreev level scattering mechanism reduces the vibronic
population to the ground state with the corresponding decay of
$I_{DC}$.

In the rotating wave approximation we ignore the possibility of
electronic scattering from the lower to the upper branch through the
emission channel. This process can occur if the time spent on
resonance, $\delta t$, is short in which case the uncertainty
principle dictates that scattering can occur not only through the
absorption but also through the non-energy conserving emission
channel. By considering also this channel for inter-Andreev level
transition, complete ground state cooling is not achievable (even in
the limit $Q\rightarrow \infty$) as the emission process will
continuously heat the oscillating nanowire. To analyze this we
consider \eqref{probs} in more detail.

Without the benefit of the RWA approximation we need to consider all
four off-diagonal terms in \eqref{probs} in order to fully describe
the dynamics of the scattering mechanism. From the form of
\eqref{probampl1} one can see that e.g. the term proportional to
$\sqrt{n-1}$ corresponds to the scattering from the lower electronic
branch in the state $n-2$ to the upper electronic branch in the state
$n-1$ through the emission of a vibron. Thus, this rate of scattering
depends on the probability of the occupancy of the harmonic oscillator
state $n-2$ and is proportional to $n-1$ as compared to the absorption
channel which, as shown above, depends on the probability of the
population of the state $n$ and scales linearly with $n$. To take
these effects into account we rewrite the unitary scattering matrix
\eqref{Smatrix},
%---------------
\begin{gather}
\hat{S}_2=\begin{pmatrix}\kappa_1'(\hat{n})& i\frac{\nu_1(\hat{n})}{\sqrt{\hat{n}+1}}\hat{b}+i\frac{\chi_1(\hat{n})}{\sqrt{\hat{n}}}\hat{b}^{\dagger}\\ i\hat{b}^{\dagger}\frac{\nu_2(\hat{n}+1)}{\sqrt{\hat{n}+1}}+i\hat{b}\frac{\chi_2(\hat{n}-1)}{\sqrt{\hat{n}}} &\kappa_2(\hat{n})\end{pmatrix}\notag\\
\vert\kappa_i'(n)\vert^2=\vert\kappa_i(n)\vert^2-\vert\chi_i(n)\vert^2\notag\\
\label{newS}
\vert\kappa_i'(n)\vert^2+\vert\nu_i(n)\vert^2+\vert\chi_i(n)\vert^2=1\,.
\end{gather}
%--------------

In \eqref{newS} $\vert\kappa_2'(n)\vert^2,\vert\nu_2(n)\vert^2$ have
the same meaning as before, i.e. they are respectively the probability
to stay or scatter (through the absorption channel) from the lower
electronic branch with oscillator in the state $n$. Similarly
$\vert\chi_2(n)\vert^2$ gives the probability that the system scatters
from the lower to the upper electronic branch through the emission
channel (the subscripts $i=1$ correspond to scattering from the upper
to the lower electronic branch). By construction, these rates are,
%----------------
\begin{gather}
\vert\nu_2(n)\vert^2=\vert\nu_1(n-1)\vert^2\notag\\
\vert\chi_1(n)\vert^2=\vert\chi_2(n-1)\vert^2\notag\\
\frac{\vert\chi_2(n-2)\vert^2}{\vert\nu_2(n)\vert^2}\frac{n}{n-1}\simeq 0.03\notag\,,
\end{gather}
%--------------------
where $\vert\nu_2(n)\vert^2=n\pi\Gamma^2$ and the coefficient of
$n/(n-1)$ in the last line was found from numerical analysis as
discussed above.

To analyze the limiting value of the stationary distribution of the
vibrational modes in the non-RWA formalism we consider the evolution
of the mechanical sub-system with $Q\rightarrow \infty$. Proceeding as
above we evaluate \eqref{rhoevol} with $\hat{S}=\hat{S}_2$ and
$\gamma=0$. From this we find that the distribution of the mechanical
modes after one period reads,
%-----------------
\begin{align}
\label{finalPemission}
P_n^{f}=P_{n+1}^{in}\vert\nu_2(n+1)\vert^2+&P_n^{in}\vert\kappa_2'(n)\vert^2+\notag\\
&P_{n-1}^{in}\vert\chi_2(n-1)\vert^2\,.
\end{align}
%------------------
Equation \eqref{finalPemission} dictates that complete ground state
cooling of the mechanical sub-system is not achievable as the term
proportional to $\vert\chi_2(n-1)\vert^2$ acts to shift the
population to higher vibrational modes and will thus compete with the
term proportional to $\vert\nu_2(n+1)$ which acts to lower
these. Considering however, that the rate of absorption is much
greater than the rate of emission we find that the stationary solution
of \eqref{finalPemission} corresponds to a population of the
vibrational modes corresponding to $\langle n\rangle=0.03$ which is in
accordance with the perturbative solution,
%----------------------
\begin{equation}
\langle n\rangle\simeq
\frac{\vert\chi_2(0)\vert^2}{\vert\nu_2(1)\vert^2}=0.03\,.
\end{equation}
%-----------------------
Thus, we find that by including not only the absorption but also the
emission channel for inter-Andreev scattering the best theoretically
achievable level of cooling is given by the ratio of these two
rates. The corresponding current with this population of the
vibrational modes would be $I_{DC}=$\unit[1.1]{pA}.

\section{\label{Appendcurr} Environmental correction to the DC current}
%%%%%%%%%%%%%%%%%%%%%%%%%%%%%%%%%%%%%%%%%%%%%%%%
%%%%%%%%%%%%%%%%%%%%%%%%%%%%%%%%%%%%%%%%%%%%%%%%
As discussed in the main text the DC current through the nanowire
scales with the average vibron population, $\langle n\rangle$. In
\eqref{current} this is shown for the case when
$P=\sum_n\vert\nu_2(n)\vert^2P_n^{stat}$. We note here that this
expression is not completely valid due to the influence of the
external environment. To understand why this is the case we have to
consider that the stationary distribution $P_n^{stat}$ gives the
vibronic population which does not change over one full period. Over
this interval of time, the mechanical sub-system changes in two ways;
by temporal evolution and through magnetic-field induced
scattering. The value of $P_n^{stat}$ depends on both of these
processes. This means that in order to accurately evaluate the total
probability of scattering of the mechanical sub-system we first have
to consider the effects of the environment on the evolution of
$P_n^{stat}$ between $0\leq t\leq T_V/2$ and calculate the total
probability of scattering from this quantity. Performing this analysis
we find that the accurate probability of scattering reads,
%------------------------
\begin{align}
\label{scattcorr}
P=\pi\Gamma^2\left[\langle n\rangle-\frac{T_V}{2Q} \left(\langle n\rangle-n_B\right)\right]\,,
\end{align}
%-----------------------
which is again linearly dependent on $\langle n\rangle$ but also
includes a correction term from the environment. 

A numerical analysis of the DC current as a function of $Q$ has been
performed using both \eqref{current} and \eqref{scattcorr}, with
little or no discrepancy between the two results. To understand why
the corrections in \eqref{scattcorr} does not change $I_{DC}$ we
analyze separately the limit of low/high damping which both give
$P=\pi\Gamma^2\langle n\rangle$. In the limit of high quality factor,
the corrections to $P_n^{stat}$ from the thermal interactions are very
small $\propto 1/Q$, hence they do not influence the stationary
mechanical distribution over the short time-span $T_V/2$ to any large
extent.  On the other hand, in the limit of low quality factor the
stationary distribution of the mechanical modes is given by the
thermal distribution, hence the correction to the total scattering
probability in \eqref{scattcorr} are identically zero. For
intermediate values of $Q$ these competing processes cancel out and
the current over the junction depends only on the average vibron
population in the stationary regime as given by equation
\eqref{current}.

%--------bibliography--------------------------
\bibliography{Paper}

\begin{thebibliography}{10}

\bibitem{Lassagne2008}
B.~Lassagne, D.~Garcia-Sanchez, A.~Aguasca, and A.~Bachtold,
\newblock Nano Lett. {\bf 8}, 3735 (2008).

\bibitem{Jensen2008}
K.~Jensen, K.~Kim, and A.~Zettl,
\newblock Nat. Nanotechnol. {\bf 3}, 533 (2008).

\bibitem{Naik2009}
A.K. Naik, M.S. Hanary, W.K. Hiebert, X.L. Feng, and M.L. Roukes,
\newblock Nature Nanotechnol. {\bf 4}, 445 (2009).

\bibitem{LaHaye}
M.D. LaHaye, O.~Buu, B.~Camarota, and K.C. Schwab,
\newblock Science {\bf 304}, 74 (2004).

\bibitem{Etaki2008}
S.~Etaki, M.~Poot, I.~Mahboob, K.~Onomitsu, H.~Yamaguchi, and H.S.J. van der Zant,
\newblock Nat. Phys. {\bf 4}, 785 (2008).

\bibitem{Hertzberg2010}
J.~B. Hertzberg, T.~Rocheleau, T.~Ndukum, M.~Savva, A.A. Clerk, and K.C. Schwab,
\newblock Nat. Phys. {\bf 6}, 213 (2010).

\bibitem{Schwab2005}
K.C. Schwab and M.L. Roukes,
\newblock Phys. Today {\bf 58}, 36 (2005).

\bibitem{Blencowe2004}
M.~Blencowe,
\newblock Phys. Rep. {\bf 395}, 159 (2004).

\bibitem{OConnel2010}
A.D. O'Connell, M.~Hofheinz, M.~Ansmann, R.C. Bialczak, M.~Lenander, E.~Lucero, M.~Neeley, D.~Sank, H.~Wang, M.~Weides, J.~Wenner, J.M. Martinis, and A.N. Cleland,
\newblock Nature 

\newblock http://dx.doi.org/10.1038/nature08967 (2010).

\bibitem{Rocheleau2009}
T.~Rocheleau, T.~Ndukum, C.~Macklin, J.B. Hertzberg, A.A. Clerk, and K.C. Schwab,
\newblock Nature {\bf 463}, 72 (2010).

\bibitem{Naik}
A.~Naik, O.~Buu, M.D. LaHaye, A.D. Armour, A.A. Clerk, M.P. Blencowe, and K.C. Schwab,
\newblock Nature {\bf 443}, 193 (2006).

\bibitem{Martin2004}
I.~Martin, A.~Shnirman, L.~Tian, and P.~Zoller,
\newblock Phys. Rev. B {\bf 69}, 125339 (2004).

\bibitem{Zippilli2009}
S.~Zippilli, G.~Morigi, and A.~Bachtold,
\newblock Phys. Rev. Lett. {\bf 102}, 096804 (2009).

\bibitem{WilsonRae2004}
I.~Wilson-Rae, P.~Zoller, and A.~Imamo\={g}lu,
\newblock Phys. Rev. Lett. {\bf 92}, 075507 (2004).

\bibitem{Ouyang2009}
S.H. Ouyang, J.Q. You, and F.~Nori,
\newblock Phys. Rev. B {\bf 79}, 075304 (2009).

\bibitem{Zippilli2010}
S.~Zippilli, A.~Bachtold, and G.~Morigi,
\newblock arXiv:1003.3816.

\bibitem{Sonne2010}
G.~Sonne, M.E.~Pe\~{n}a-Aza, L.Y. Gorelik, R.I. Shekhter, and M.~Jonson,
\newblock arXiv:1002.1207.

\bibitem{Bagwell}
P.F. Bagwell,
\newblock Phys. Rev. B {\bf 46}, 12573 (1992).

\bibitem{Beenakker1991}
C.W.J. Beenakker,
\newblock Phys. Rev. Lett. {\bf 67}, 3836 (1991).

\bibitem{Gorelik1995}
L.Y. Gorelik, V.S. Shumeiko, R.I. Shekhter, G.~Wendin, and M.~Jonson,
\newblock Phys. Rev. Lett. {\bf 75}, 1162 (1995).

\bibitem{Kong2001}
J.~Kong, E.~Yenilmez, T.W. Tombler, W.~Kim, H.~Dai, R.B. Laughlin, L.~Liu, S.~Jayanthi, and S.Y. Wu,
\newblock Phys. Rev. Lett. {\bf 87}, 106801 (2001).

\bibitem{Gorelik1998}
L.Y. Gorelik, N.I. Lundin, V.S. Shumeiko, R.I. Shekhter, and M.~Jonson,
\newblock Phys. Rev. Lett. {\bf 81}, 2538 (1998).

\bibitem{Huttel2009}
A.K. H\"{u}ttel, G.A. Steele, B.~Witkamp, M.~Poot, L.P. Kouwenhoven, and H.S.J. van der Zant,
\newblock Nano Lett. {\bf 9}, 2547 (2009).

\end{thebibliography}

\end{document}